%% file: main.tex
\documentclass[10pt,twocolumn,letterpaper]{article}

\usepackage{cvpr}              
\usepackage[accsupp]{axessibility}
\input{preamble}

\definecolor{cvprblue}{rgb}{0.21,0.49,0.74}
\usepackage[pagebackref,breaklinks,colorlinks,allcolors=cvprblue]{hyperref}

\def\paperID{*****} 
\def\confName{CVPR}
\def\confYear{2025}

\title{\textsc{Rewind}: Real-Time Egocentric Whole-Body Motion Diffusion \\with Exemplar-Based Identity Conditioning}

\author{Jihyun Lee$^{1,2}$
\quad
Weipeng Xu$^{1}$
\quad
Alexander Richard$^{1}$
\quad
Shih-En Wei$^{1}$
\quad
Shunsuke Saito$^{1}$
\\
Shaojie Bai$^{1}$
\quad
Te-Li Wang$^{1}$
\quad
Minhyuk Sung$^{2}$
\quad
Tae-Kyun Kim$^{2,3}$
\quad
Jason Saragih$^{1}$\\\\
$^1$ Codec Avatars Lab, Meta \quad $^2$ KAIST \quad $^3$ Imperial College London\\
\small{\url{https://jyunlee.github.io/projects/rewind}}
}
\vspace{-2\baselineskip}

\begin{document}

\twocolumn[{%
\maketitle

\renewcommand\twocolumn[1][]{#1}%
\maketitle
\begin{center}
\centering
\vspace{-0.3\baselineskip}
\captionsetup{type=figure}
\includegraphics[width=0.95\textwidth]{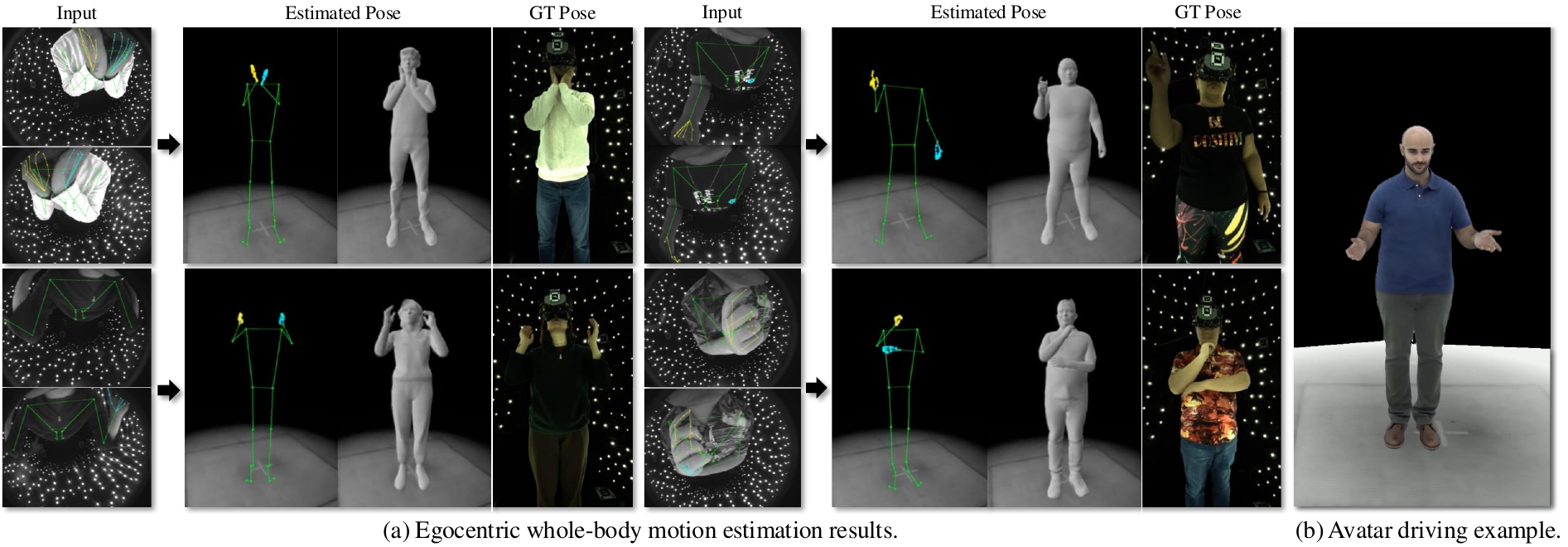} 
\vspace{-0.4\baselineskip}
\caption{\textbf{Real-time and high-fidelity whole-body motion estimation from stereo egocentric images.}
We propose \textsc{Rewind}, a novel egocentric image-conditioned diffusion model for high-quality 3D whole-body motion estimation.
\textsc{Rewind} is real-time, causal, and generalizable to unseen motion lengths, making it seamlessly applicable for driving photorealistic avatars or meshes.
Please refer to the supplementary video, which demonstrates that \textsc{Rewind} estimates significantly more plausible motions compared to existing methods.
}
\vspace{0.3\baselineskip}
\label{fig:teaser_image}
\end{center}
}]

\input{sec/0_abstract}    
\input{sec/1_intro}
\input{sec/2_related_work}
\input{sec/3_method}

\input{sec/4_experiments}
\input{sec/5_conclusion}

{
    \small
    \bibliographystyle{ieeenat_fullname}
    \bibliography{main}
}

\input{sec/6_suppl}

\end{document}

%% file: preamble.tex
\usepackage{graphicx}
\usepackage{amsmath}
\usepackage{amssymb}
\usepackage{booktabs}
\usepackage{multirow}
\usepackage{booktabs}
\usepackage{tabularx}
\usepackage{makecell}
\usepackage{multirow}
\usepackage{makecell}
\usepackage{marvosym}
\usepackage{capt-of}
\usepackage{comment}
\usepackage{multicol}
\usepackage{algorithm}
\usepackage{algcompatible}
\usepackage{algorithmicx}
\usepackage{algpseudocode}
\usepackage{subcaption}
\usepackage[accsupp]{axessibility}
\usepackage[dvipsnames]{xcolor}

\newcolumntype{Y}{>{\centering\arraybackslash}X}

\usepackage{stackengine}

\algnewcommand{\LeftComment}[1]{\Statex \(\triangleright\) #1}

\usepackage[dvipsnames]{xcolor}

\newcommand{\keyword}[1]{{\noindent\textbf{#1}}}

\definecolor{color_1}{RGB}{255,0,128}
\definecolor{color_2}{RGB}{0,128,128}
\definecolor{color_3}{RGB}{0,128,0}
\definecolor{color_4}{RGB}{128,0,0}

%% file: sec/0_abstract.tex
\begin{abstract}
\label{sec:abstract}
\vspace{-1.5\baselineskip}

We present \textsc{Rewind} (\textbf{R}eal-Time \textbf{E}gocentric \textbf{W}hole-Body Mot\textbf{i}o\textbf{n} \textbf{D}iffusion), a one-step diffusion model for real-time, high-fidelity human motion estimation from egocentric image inputs.
While an existing method for egocentric whole-body (i.e., body and hands) motion estimation is non-real-time and acausal due to diffusion-based iterative motion refinement to capture correlations between body and hand poses, \textsc{Rewind} operates in a fully causal and real-time manner.
To enable real-time inference, we introduce (1) cascaded body-hand denoising diffusion, which effectively models the correlation between egocentric body and hand motions in a fast, feed-forward manner, and (2) diffusion distillation, which enables high-quality motion estimation with a single denoising step.
Our denoising diffusion model is based on a modified Transformer architecture, designed to causally model output motions while enhancing generalizability to unseen motion lengths.
Additionally, \textsc{Rewind} optionally supports identity-conditioned motion estimation when identity prior is available.
To this end, we propose a novel identity conditioning method based on a small set of pose exemplars of the target identity, which further enhances motion estimation quality.
Through extensive experiments, we demonstrate that \textsc{Rewind} significantly outperforms the existing baselines both with and without exemplar-based identity conditioning.

\vspace{-\baselineskip}

\end{abstract}

\vspace{-0.5\baselineskip}

%% file: sec/1_intro.tex
\section{Introduction}
\label{sec:intro}
\vspace{-0.2\baselineskip}
%
%
Egocentric human motion estimation is essential for delivering immersive and realistic experiences in AR/VR applications, such as gaming and telepresence.
For instance, imagine engaging in a conversation with your friend in a virtual environment.
The quality of estimated whole-body (i.e., body and hands) motion is crucial in creating a realistic experience in interpersonal communication, while subtle pose changes (e.g., finger poses in Fig.~\textcolor{RoyalBlue}{1b}) can significantly impact the intended message.

Developing a live-drivable method that enables accurate and realistic egocentric whole-body motion estimation is thus essential.
However, existing egocentric pose or motion estimation methods fall short in achieving the accuracy and speed needed for highly realistic VR/AR experiences.
They typically focus on tracking \emph{body-only} motions~\cite{yang2024egoposeformer, cuevas2024simpleego, akada2022unrealego, wang2023scene, wang2021estimating, akada20243d, kang2024attention}, neglecting the importance of hands in fully capturing the intricacies of
human motions~\cite{wang2024egocentric}.
Directly extending the existing body-only estimation methods for whole-body estimation yields suboptimal results, as body and hands significantly differ in scale in both input images and output motions~\cite{rong2021frankmocap, choutas2020monocular,zhou2021monocular,feng2021collaborative,moon2022accurate,zhang2023pymaf}.

To address this challenge, EgoWholeMocap~\cite{wang2024egocentric}, the first method for whole-body motion estimation from egocentric images, proposes to leverage \emph{specialist models} for body and hands.
It first performs per-frame pose estimation for body and hands separately, and then refines the output poses using an unconditional whole-body motion prior to model correlations between different body parts.
While this approach improves whole-body motion estimation performance, it is non-real-time and acausal (i.e., depends on future information) due to iterative refinement steps using an acausal diffusion-based motion prior.
Thus, it cannot be used for real-time egocentric motion tracking applications.

In this work, we introduce \textsc{Rewind} (\textbf{R}eal-Time \textbf{E}gocentric \textbf{W}hole-Body Mot\textbf{i}o\textbf{n} \textbf{D}iffusion), a one-step diffusion model for real-time, high-fidelity human motion estimation from egocentric image inputs.  
To achieve both fast inference speed and high whole-body motion accuracy, we first introduce \textbf{cascaded body-hand denoising diffusion (Sec.~\ref{subsubsec:decomposed_body_hand_diffusion})}, where body motion is sampled first and then hand motion is sampled \emph{conditioned} on the previously sampled upper body motion.
This cascading scheme approximately models the correlation between body and hands in a fast, feed-forward manner (cf. iterative whole-body refinement used in EgoWholeMocap~\cite{wang2024egocentric}) while still inheriting the advantages of specialized body and hand estimation (e.g., effectively handling domain differences).
We further argue that this approach is particularly effective for our targeted egocentric image inputs, where hands are often placed outside the field of view or occluded. 
As hands and upper body poses are known to have meaningful correlations~\cite{ng2021body2hands}, conditioning on estimated upper body motion -- often provided with more reliable input egocentric observations (e.g., Fig.~\textcolor{RoyalBlue}{1a}) -- can effectively reduce hand motion ambiguities.


We build these specialist denoising diffusion models based on \textbf{causal relative-temporal Transformer (Sec.~\ref{subsubsec:network_relative_temporal_transformer})}, which is fully causal and generalizable to unseen motion lengths. 
We use windowed relative-temporal attention to learn temporal motion features that are invariant to the total sequence length or absolute timesteps, in contrast to motion diffusion models based on vanilla attention (e.g., motion prior used in EgoWholeMocap~\cite{wang2024egocentric}).
During network training, we employ \textbf{diffusion distillation (Sec.~\ref{subsubsec:diffusion_distillation})} to enable real-time inference ($>$ 30 FPS) with \emph{a single denoising step}, while preserving high output motion quality.

Not only introducing an effective real-time framework, we take a step further and explore \emph{identity}-aware motion estimation to further enhance output quality when an additional identity prior is available.
To this end, we propose novel \textbf{exemplar-based identity conditioning (Sec.~\ref{subsec:personalized_motion})}, where motion estimation is conditioned on the target identity parameterized by a small set of pose exemplars.
While this identity parameterization has not yet been considered in existing works on human pose or motion estimation, we empirically find it to be the most effective compared to widely used identity parameterizations (e.g., height, bone lengths, shape parameters).
%
%
In the experiments, \textsc{Rewind} achieves state-of-the-art whole-body motion estimation results, both with and without additional identity priors.
Please also refer to our supplementary video, where \textsc{Rewind} estimates significantly more plausible motions than the baselines~\cite{yang2024egoposeformer, wang2024egocentric}, even from challenging egocentric input observations (e.g., occluded or truncated views).
%

\vspace{-1.2\baselineskip}
%
%

%% file: sec/2_related_work.tex
\section{Related Work}
\label{sec:related_work}

\vspace{-0.2\baselineskip}

\subsection{Egocentric Body Pose Estimation}
\label{subsec:egocentric_body_pose_estimation}
Recently, various egocentric body pose or motion estimation methods have been proposed for different input domains (e.g., sensors~\cite{jiang2022avatarposer,jiang2023egoposer,du2023avatars} or images~\cite{li2023ego,yi2024estimating,wang2024egocentric, yang2024egoposeformer}).
Here, we focus on existing methods for estimating the pose of a head-mounted device wearer from image inputs, which are most relevant to our work.
These methods can be broadly categorized into two groups based on whether they utilize \textit{front-facing} or \textit{down-facing} egocentric cameras.
Methods using \textit{front-facing} egocentric cameras~\cite{li2023ego,yi2024estimating,ng2020you2me,ma2024nymeria,li2024egogen,guzov2024hmd} assume that the wearer's body is not visible from the input viewpoint.
Thus, they formulate the problem as a motion generation or inpainting task, conditioned on head-mounted camera poses~\cite{li2023ego, yi2024estimating}, hand poses~\cite{yi2024estimating}, or the body poses of social interactees~\cite{ng2020you2me}.
On the other hand, methods using \textit{down-facing} egocentric cameras~\cite{wang2024egocentric, yang2024egoposeformer, cuevas2024simpleego, akada2022unrealego, wang2023scene, wang2021estimating, akada20243d, kang2024attention} focus on recovering 3D body poses from visual observations. However, they still suffer from self-occlusions and truncated views caused by the egocentric viewpoint.
To address these challenges, some methods incorporate motion priors~\cite{wang2024egocentric, wang2021estimating} or scene information~\cite{wang2023scene, akada20243d} to reduce pose ambiguities, while others propose novel network architectures to better handle uncertainty~\cite{kang2024attention, cuevas2024simpleego}.
In this work, we focus on egocentric motion estimation using stereo down-facing cameras.
Unlike most existing methods, which estimate body-only poses~\cite{yang2024egoposeformer, cuevas2024simpleego, akada2022unrealego, wang2023scene, wang2021estimating, akada20243d, kang2024attention}, we aim to estimate whole-body poses (i.e., body and hands) for more comprehensive motion modeling.

%
\subsection{Whole-Body Pose Estimation}
\label{subsec:whole_body_pose_estimation}
Whole-body pose or motion estimation aims to jointly predict the poses of body and hands.
The main technical challenge lies in the scale and pose distribution differences between different body parts.
To address this, most existing works~\cite{rong2021frankmocap, choutas2020monocular,zhou2021monocular,feng2021collaborative,moon2022accurate,zhang2023pymaf} use separate models to predict each body part and merge the results, often with an optional integration network~\cite{feng2021collaborative,zhang2023pymaf} or post-processing~\cite{rong2021frankmocap} to improve alignment between the body parts.
However, these methods primarily focus on exocentric image inputs, leaving egocentric whole-body pose estimation largely unexplored.
Recently, EgoWholeMocap~\cite{wang2024egocentric} introduced the \emph{first} whole-body pose estimation method for egocentric image inputs, based on separate body and hand pose estimation with diffusion-based motion refinement.
While EgoWholeMocap employs an \emph{unconditional} whole-body motion diffusion prior for post-processing, we directly train a motion diffusion model \emph{conditioned} on egocentric inputs to predict motion that is more coherent with the input observation.
%

\subsection{Motion Diffusion Models}
\label{subsec:motion_diffusion_models}
%
%
We review existing motion diffusion models that model \emph{arbitrarily long motion} and \emph{identity-conditioned motion}, which are two key objectives of our work.
Since these challenges remain underexplored in \emph{egocentric} motion estimation, we primarily discuss prior work on \emph{unconditional} or \emph{text-conditional} motion generation.

\noindent \textbf{Arbitrarily long motion.}
Some works propose motion diffusion models that can generalize to motions longer than training instances~\cite{shafir2023human, petrovich2024multi, zhang2023diffcollage, barquero2024seamless}.
For example, DoubleTake~\cite{shafir2023human}, STMC~\cite{petrovich2024multi}, and DiffCollage~\cite{zhang2023diffcollage} propose generating multiple motion segments, each with a temporal length within the training distribution, and then applying a special sampling mechanism to smoothly combine them into a longer motion.
However, these methods rely on future information for motion composition. 
The most related work to ours is FlowMDM~\cite{barquero2024seamless}, which introduces a novel Transformer-based architecture~\cite{vaswani2017attention} using relative positional encoding~\cite{su2024roformer} to improve temporal extrapolation.
However, it still relies on future information and partially incorporates absolute positional encoding~\cite{vaswani2017attention}, which limits its temporal generalization capability.
In this work, we propose utilizing relative positional encoding~\cite{su2024roformer} similar to FlowMDM, but we \emph{completely} eliminate dependencies on (1) absolute frame timesteps to extract motion features invariant to sequence length, and (2) future information to make it more suitable for real-time applications (Sec.~\ref{subsubsec:network_relative_temporal_transformer}).

\noindent \textbf{Identity-conditioned motion.}
A few recently proposed methods focus on identity-conditioned motion generation~\cite{tripathi2025humos,xue2024shape}.
HUMOS~\cite{tripathi2025humos,xue2024shape} conditions the motion diffusion model on SMPL~\cite{loper2015smpl} shape parameters.
Due to the lack of datasets with paired motion and identity annotations~\cite{tripathi2025humos}, it proposes a novel loss function to learn identity-specific motions from unpaired training data. 
SMD~\cite{xue2024shape} introduces a spectral feature encoder to integrate the template mesh of the target identity into the motion diffusion model.
Inspired by these motion diffusion models proposed for unconditional or text-conditional motion generation, we introduce the first method for identity-conditioned \emph{egocentric} motion estimation.

\vspace{-0.2\baselineskip}

%% file: sec/3_method.tex
\section{Egocentric Whole-Body Motion Estimation}
\label{sec:method}

\begin{figure*}[!t]
\begin{center}
\includegraphics[width=0.92\textwidth]{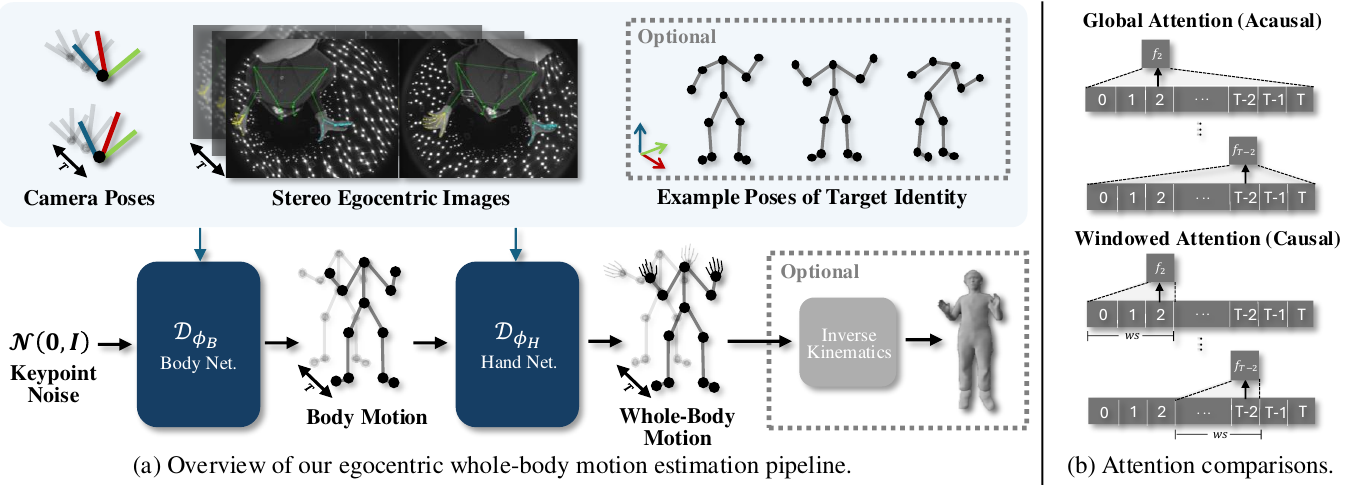} 
\vspace{-0.6\baselineskip}
\caption{\textbf{(a) Pipeline overview.}
Given a sequence of stereo egocentric images and camera poses, our diffusion model first estimates 3D body motion and then estimates 3D hand motion conditioned on the 3D upper body motion. 
Our motion estimation can be optionally conditioned on the exemplar-based identity prior when available (Sec.~\ref{subsec:personalized_motion}). 
Through an optional inverse kinematics step (refer to the supplementary for details), our tracking results can be used to drive meshes or photorealistic avatars.
\textbf{(b) Attention comparisons.}
Compared to vanilla self-attention (i.e., acausal, global attention) commonly used in existing works, the proposed causal windowed attention conditioned on relative timesteps enhances generalization to unseen motion lengths (Sec.~\ref{subsubsec:network_relative_temporal_transformer}).
%
}
\vspace{-1.7\baselineskip}
\label{fig:pipeline_overview}
\end{center}
\end{figure*}

Our goal is to estimate first-person whole-body motion from egocentric image inputs in real time.
Motivated by existing image-based pose or motion estimation methods that demonstrate that diffusion models~\cite{ho2020denoising, song2021denoising} are effective at handling occluded or out-of-view body observations~\cite{stathopoulos2024score, holmquist2023diffpose, zhou2023diff3dhpe, gong2023diffpose, feng2023diffpose, zhang2023probabilistic, yi2024estimating, wang2024egocentric}, we propose a diffusion-based approach. 
Formally, our denoising diffusion network models whole-body motion conditioned on input egocentric observations:

\vspace{-0.5\baselineskip}
\begin{equation}
p_{\phi}\,(\mathbf{J}^{1:T}\, |\, \Phi^{1:T}),
\label{eq:conditional_whole_body_dist}
\end{equation}

\noindent where $p_{\phi}$ denotes the model distribution parameterized by the diffusion network weights $\phi$.
$\mathbf{J}^{1:T}$ represents a sequence of whole-body poses, and $\Phi^{1:T}$ denotes a sequence of input egocentric observations over $T$ frames. 
At each timestep $t \in (1, T)$, a whole-body pose $\mathbf{J}^{t}$ is represented by $N_{\mathbf{J}}$ number of 3D keypoints, and an egocentric observation $\Phi^{t}$ consists of stereo egocentric images and camera poses:
\begin{equation}
\Phi^{t} = [\mathbf{I}_{\textit{L}},\, \mathbf{I}_{\textit{R}}, \mathbf{C}_{\textit{L}},\, \mathbf{C}_{\textit{R}}].
\label{eq:egocentric_observation}
\end{equation}
\noindent $\mathbf{I}_{v \in \{\textit{L},\, \textit{R}\}} \in \mathbb{R}^{C \times W \times H}$ is an egocentric image captured from the viewpoint $v$, and $\mathbf{C}_{v \in \{L,\, R\}} = [\mathbf{R}_{v} | \mathbf{t}_{v}] \in \mathbb{R}^{3 \times 4}$ is the corresponding camera pose, with camera rotation $\mathbf{R}_{v} \in \mathbb{R}^{3 \times 3}$ and translation $\mathbf{t}_{v} \in \mathbb{R}^{3 \times 1}$.
Note that SLAM systems in recent head-mounted devices~\cite{engel2023project} achieve millimeter-level accuracy~\cite{yi2024estimating}, thus camera poses are considered as additional inputs in recent egocentric tracking methods~\cite{yi2024estimating, luo2024real, guzov2024hmd}. 
In the following subsections, we discuss each component of our method, designed to achieve real-time, fully causal whole-body motion estimation.
%
%

\subsection{Cascaded Body-Hand Denoising Diffusion}
\label{subsubsec:decomposed_body_hand_diffusion}
Whole-body motion estimation is challenging due to the scale and pose distribution differences between body and hands~\cite{rong2021frankmocap, choutas2020monocular, zhou2021monocular, feng2021collaborative, moon2022accurate, zhang2023pymaf}.
To address this, the current state-of-the-art method for egocentric whole-body motion estimation (EgoWholeMocap~\cite{wang2024egocentric}) employs specialist models for body and hand pose estimation to handle domain differences, along with output refinement using an unconditional motion diffusion prior to model correlations between different body parts.
However, we argue that this approach may be suboptimal, because (1) the additional refinement steps slow down inference speed, and (2) the use of an \emph{unconditional} motion prior is less effective for predicting motions highly coherent with the input image observations.
To address these, we propose \emph{cascaded body-hand denoising diffusion}, a crucial component that enhances both the accuracy and efficiency of egocentric whole-body motion estimation.

In a nutshell, our idea is to first estimate a body motion, and then condition the subsequent hand motion estimation on the estimated 3D upper body motion. 
This was inspired by existing work~\cite{ng2021body2hands} that demonstrated a meaningful correlation between 3D upper body and hand poses.
Note that our cascading approach enables the fast, feed-forward capture of the approximated correlation between body and hands (cf. iterative whole-body refinement in EgoWholeMocap~\cite{wang2024egocentric}), while still benefiting from specialized body and hand estimation to effectively handle domain differences.
We also argue that this approach is particularly effective for egocentric hand estimation, where input hand observations are often highly ambiguous (e.g., hands are frequently placed outside of the field of view or occluded by other body parts as shown in Fig.~\ref{fig:teaser_image}).
By conditioning the output hand motion on the estimated 3D upper body motion, which is often more reliably observed in the input egocentric views, we can effectively reduce ambiguity in hand estimation.

Formally, we reformulate the egocentric-conditioned whole-body motion distribution in Eq.~\ref{eq:conditional_whole_body_dist} as:
\vspace{0.1\baselineskip}
\begin{equation}\small{
p_{\phi}\,(\mathbf{J}^{1:T}\, |\, \Phi^{1:T}) \approx p_{\phi_{\textit{B}}}\,(\mathbf{J}^{1:T}_{\textit{B}}\, |\, \Phi^{1:T})\, p_{\phi_{\textit{H}}}\,(\mathbf{J}^{1:T}_{\textit{H}}\, |\, \mathbf{J}^{1:T}_{\textit{B}_{\textit{upper}}},\, \Phi^{1:T}),}
\label{eq:body_hand_decomposition}
\end{equation}
\noindent where the subscripts $B$, $B_{\textit{upper}}$, and $H$ represent the body, upper body, and hands, respectively.
During training, we separately train body and hand specialist models to learn $p_{\phi_{\textit{B}}}\,(\mathbf{J}^{1:T}_{\textit{B}}\, |\, \Phi^{1:T})$ and $p_{\phi_{\textit{H}}}\,(\mathbf{J}^{1:T}_{\textit{H}}\, |\, \mathbf{J}^{1:T}_{\textit{B}_{\textit{upper}}},\, \Phi^{1:T})$, respectively.
During inference, we simply sample from each of the learned distributions in a cascaded manner.
In the experiments (Sec.~\ref{sec:experiments}), we empirically demonstrate that this cascaded approach outperforms (1) a method that estimates body and hands with specialist models followed by iterative whole-body refinement (EgoWholeMocap~\cite{wang2024egocentric}), and (2) a method tha estimates body and hands in a joint, parallel manner.

%
%

\subsection{Causal Relative-Temporal Transformer}
\label{subsubsec:network_relative_temporal_transformer}
We now describe our network architecture design for the specialist models for body and hands.
Recent motion diffusion models have demonstrated that Transformer encoder-based architectures are highly effective for learning motion distributions and have become the dominant choice in the field (e.g., \cite{tevet2022human, shafir2023human, tripathi2025humos, wang2024egocentric, barquero2024seamless, tseng2023edge, zhang2024motiondiffuse}).
However, these models typically generate fixed-length motions using vanilla self-attention with absolute timestep encoding, which limits their ability to generalize to motion lengths unseen during training.
To address this, several methods have been proposed for diffusion-based long motion generation or composition~\cite{barquero2024seamless,zhang2023diffcollage,shafir2023human,petrovich2024multi}, but they rely on future information for temporal extrapolation, as discussed in Sec.~\ref{subsec:motion_diffusion_models}.

In this work, we introduce the \emph{causal relative-temporal Transformer}, a modified Transformer encoder-based architecture that learns temporal features invariant to total motion length or future frames, making it fully causal and inherently generalizable to arbitrary motion lengths.
In a nutshell, our key idea is to adopt Rotary Positional Encoding (RoPE)~\cite{su2024roformer} to condition attention scores on \emph{relative} temporal distances between input tokens while restricting each token's neighborhood (i.e., the domain over which self-attention is applied) to $\mathit{ws} \in \mathbb{N}$ past frames.
Formally, our self-attention function $\mathcal{A}(\cdot, \cdot, \cdot)_{j}$ for $j$-th frame given query, key value matrices is defined as:
\begin{equation}
\mathcal{A}(\mathbf{Q}, \mathbf{K}, \mathbf{V})_{j} = \frac{\sum_{i=j-\textit{ws}}^{j}\mathbf{R}_{j}\,\theta(\mathbf{q}_{j})^{\mathbf{T}}\, \mathbf{R}_{i}\,\rho(\mathbf{k}_{i})\,\mathbf{v}_{i}}{\sum^{j}_{i=j-\textit{ws}}\theta(\mathbf{q}_{j})^{\mathbf{T}}\,\rho(\mathbf{k}_{i})},
\label{eq:self_attention}
\end{equation}
\noindent where $\mathbf{Q}, \mathbf{K}, \mathbf{V} \in \mathbb{R}^{D \times T}$ are query, key, and value matrices, and $\mathbf{q}_{i}, \mathbf{k}_{i}, \mathbf{v}_{i} \in \mathbb{R}^{D}$ denote the column vectors of each matrix for timestep $i$, respectively.
$\theta(\cdot)$ and $\rho(\cdot)$ are feature projection functions (e.g., MLP).
$\mathbf{R}_{i} \in \mathrm{SO}(D)$ is a $D$-dimensional rotation matrix parameterized by timestep $i$ as proposed in \cite{su2024roformer}. 
Note that the attention score, involving the dot product between $\mathbf{R}_{j}\,\theta(\mathbf{q}_{j})$ and $\mathbf{R}_{i}\,\rho(\mathbf{k}_{i})$, depends on the relative rotation $\mathbf{R}_{i-j}$ parameterized by the \emph{relative} timestep of the $i$-th token with respect to the $j$-th token. 
Thus, the output features remain invariant to their absolute timesteps, unlike the positional encoding used in vanilla self-attention~\cite{vaswani2017attention}.
In addition, our self-attention for $j$-th frame is performed over the input frames within the temporal window $[j-\textit{ws},\, j]$.
Since the output features depend only on a fixed number of past frames, they remain invariant to the total motion length and do not rely on future information.
In the experiments (Sec.~\ref{sec:experiments}), we demonstrate the effectiveness of this design choice in comparison to other temporal model variants.

\noindent \textbf{Building body and hand specialist models.}
Using the proposed causal relative-temporal Transformer, we now discuss the details of building the denoising diffusion networks $\mathcal{D}_{\phi_{\textit{B}}}(\cdot)$ and $\mathcal{D}_{\phi_{\textit{H}}}(\cdot)$ to model the distributions $p_{\phi_{\textit{B}}}(\cdot)$ and $p_{\phi_{\textit{H}}}(\cdot)$ in Eq.~\ref{eq:body_hand_decomposition}, respectively.
Note that we use the same network design for both the body and hand specialist models, with the only difference being that the hand model takes an additional upper body conditioning input.
Thus, for simplicity, we will base our explanation on the body model and omit the body and hand subscripts ($B$ and $H$).
In overview, our network takes as inputs a sequence of egocentric observations $\Phi^{1:T}$, a sequence of diffused keypoints $\tilde{\mathbf{J}}^{1:T}_{k}$, and the corresponding diffusion time $k$, and estimates a sequence of clean keypoints $\tilde{\mathbf{J}}^{1:T}_{0}$ at diffusion time 0.
First, we extract frame-based features for the egocentric observations at each timestep $t$ by encoding (1) 2D keypoints and their uncertainty scores estimated from the images, (2) camera parameters, and (3) diffusion time.
Next, we concatenate these conditioning features to the input diffused keypoints $\tilde{\mathbf{J}}^{1:T}_{k}$ and apply graph convolutions~\cite{defferrard2016convolutional} on the human skeletal graph to extract structural features.
We then apply our causal relative-temporal transformer (Sec.~\ref{subsubsec:network_relative_temporal_transformer}) to extract temporal features, followed by a regression head to estimate the final motion.
For the diffusion formulation, we use DDPM~\cite{ho2020denoising} for training and DDIM~\cite{song2021denoising} for inference.
For additional implementation and training details (e.g., full loss functions), we refer readers to the supplementary material.

\subsection{Diffusion Distillation}
\label{subsubsec:diffusion_distillation}
While diffusion models have shown effective for human pose or motion estimation~\cite{stathopoulos2024score, holmquist2023diffpose, zhou2023diff3dhpe, gong2023diffpose, feng2023diffpose, zhang2023probabilistic, yi2024estimating, wang2024egocentric}, their inference is typically slow due to multi-step sampling.
To mitigate this, we leverage diffusion distillation~\cite{yin2024one, geng2024one, sauer2025adversarial} to improve sampling efficiency.
Specifically, we distill the original multi-step diffusion model $\mathcal{D}^{T}_{\phi}$ into a single-step lightweight model $\mathcal{D}^{S}_{\phi*}$ using Score Distillation Sampling (SDS) loss, inspired by~\cite{sauer2025adversarial, poole2023dreamfusion}.
Given the keypoints estimated by the student model $\hat{\mathbf{J}}^{1:T}_{0} \leftarrow \mathcal{D}_{\phi*}^{S}(\tilde{\mathbf{J}}^{1:T}_{K}, \Phi^{1:T}, K)$, where $K$ denotes the maximum diffusion timestep, our distillation loss is defined as:

\vspace{-\baselineskip}
\begin{equation}
\mathcal{L}_{\mathit{distill}} = ||\, \mathcal{D}_{\phi}^{T}(\mathcal{E}(\hat{\mathbf{J}}^{1:T}_{0}, k_{\mathit{small}}), \Phi^{1:T}, k_{\mathit{small}}) - \hat{\mathbf{J}}^{1:T}_{0}\, ||_{2},
\label{eq:sds}
\end{equation}

\noindent where $\mathcal{E}(\cdot)$ is a forward diffusion function~\cite{ho2020denoising, poole2023dreamfusion} that adds a small noise corresponding to diffusion timestep $k_{\textit{small}}$ to the estimated keypoints $\hat{\mathbf{J}}^{1:T}_{0}$.
Intuitively, this distillation loss encourages the student model to sample keypoints that the teacher model deems plausible when conditioned on the same egocentric observations.
Unlike the existing approach~\cite{sauer2025adversarial} that incorporates an additional adversarial loss to improve single-step sampling quality for image generation, we find that SDS loss alone is sufficient to achieve SotA results in egocentric motion estimation while achieving an inference speed of over 30 FPS.

%

\subsection{Exemplar-Based Identity Conditioning}
\label{subsec:personalized_motion}
%
While the proposed method already achieves state-of-the-art results in egocentric motion estimation, we take a step further by exploring \emph{identity}-aware motion estimation to further enhance output quality.
We hypothesize that incorporating prior information about the identity performing the motion (e.g., body shape, posture style) can help reduce motion ambiguity when such prior is additionally available.
In particular, we find that \emph{exemplar-based identity conditioning}, which conditions the output motion on \emph{a small set of example 3D poses} of the target identity, is highly effective.
This approach is inspired by recent work on representation learning for face images~\cite{bai2024universal}, which demonstrates that conditioning on a set of example images of the target identity is effective for learning identity-aware features, significantly improving final reconstruction performance (see \cite{bai2024universal} for details).
Analogously, in our work, example poses of the target identity can provide useful information about body scale and posture style of that particular person.

Formally, let $\{\mathbf{J}_{i}^{\mathcal{I}}\}_{i=1\,, ...,\, N_{O}}$ denote a set of $N_{O}$ example poses of the target identity $\mathcal{I}$ observed prior to the motion estimation phase, where $\mathbf{J}_{i}^{\mathcal{I}} \in \mathbb{R}^{N_{\mathbf{J}} \times 3}$ is represented as 3D keypoints.
This pose set can be obtained, for example, through a simple pose registration stage where we capture monocular photos of the target person performing natural motions and estimate 3D poses from these images.
In our experiments, we estimate these poses by fitting a parametric body model to 2D keypoints estimated from the input images using an off-the-shelf model (Sapiens~\cite{khirodkar2025sapiens}), along with the person's height to resolve scale ambiguity (see the supplementary material for details).
Once the example poses are registered, they can be used to enhance the quality of all subsequent motion estimation sessions for that identity.
Notably, this prior is less cumbersome to acquire than other priors (e.g., registered scene geometry) used in some of the existing egocentric motion estimation methods~\cite{wang2023scene,akada20243d} to reduce motion ambiguities.

Given a set of example poses for the target identity, we perform set encoding to extract features invariant to the order of poses.
We apply an MLP-based encoder $\gamma(\cdot)$ shared across the input poses and aggregate the resulting features using a symmetric function $\rho(\cdot)$ as follows:
\vspace{-1.1\baselineskip}

\begin{equation}
f_{\textit{ex}}^{\mathcal{I}} = \rho \, (\gamma\,(\mathbf{J}_{0}^{\mathcal{I}}),\, \gamma\,(\mathbf{J}_{1}^{\mathcal{I}}),\, ...,\, \gamma\,(\mathbf{J}_{O-1}^{\mathcal{I}}),\, \gamma(\mathbf{J}_{O}^{\mathcal{I}})).
\label{eq:anchor_pose_feature_encoding}
\vspace{-0.3\baselineskip}
\end{equation}
\noindent In practice, we instantiate $\rho(\cdot)$ as a max-pooling function. 
We finally incorporate this identity feature $f_{\textit{ex}}^{\mathcal{I}}$ into our framework using AdaIN~\cite{huang2017arbitrary}, a technique widely used for incorporating style conditions.
In Sec.~\ref{sec:experiments}, we empirically demonstrate that this exemplar-based identity prior results in greater performance improvements compared to other identity priors (e.g., shape parameters, bone lengths).
To the best of our knowledge, this is also the \emph{first} study to analyze the effectiveness of different identity priors in egocentric motion estimation.

%% file: sec/4_experiments.tex
\section{Experiments}
\label{sec:experiments}

%
%
%

\subsection{Dataset}
\label{subsec:experimental_setups}

%
Unlike exocentric (i.e., third-person view) image-based motion estimation, there had been no benchmark proposed for egocentric \emph{whole-body} motion estimation with high-quality body and hand annotations.
To address this, EgoWholeMocap~\cite{wang2024egocentric} has recently created a large-scale syntehtic dataset.
However, only their samples for training frame-based models (i.e., temporally discontinuous samples) are publicly available, limiting their use for our temporal model experiments. 
The synthetic dataset created by SimpleEgo~\cite{cuevas2024simpleego} also contains whole-body pose annotations, but it is not temporal as well.
Thus, we consider the following datasets for our experiments: (1) \emph{ColossusEgo}, a large-scale real dataset that we have newly collected, and (2) \emph{UnrealEgo}~\cite{akada2022unrealego, akada20243d}, a synthetic dataset originally proposed for egocentric body-only pose estimation but containing auxiliary hand annotations.

\begin{figure*}[!t]
\begin{center}
\includegraphics[width=0.94\textwidth]{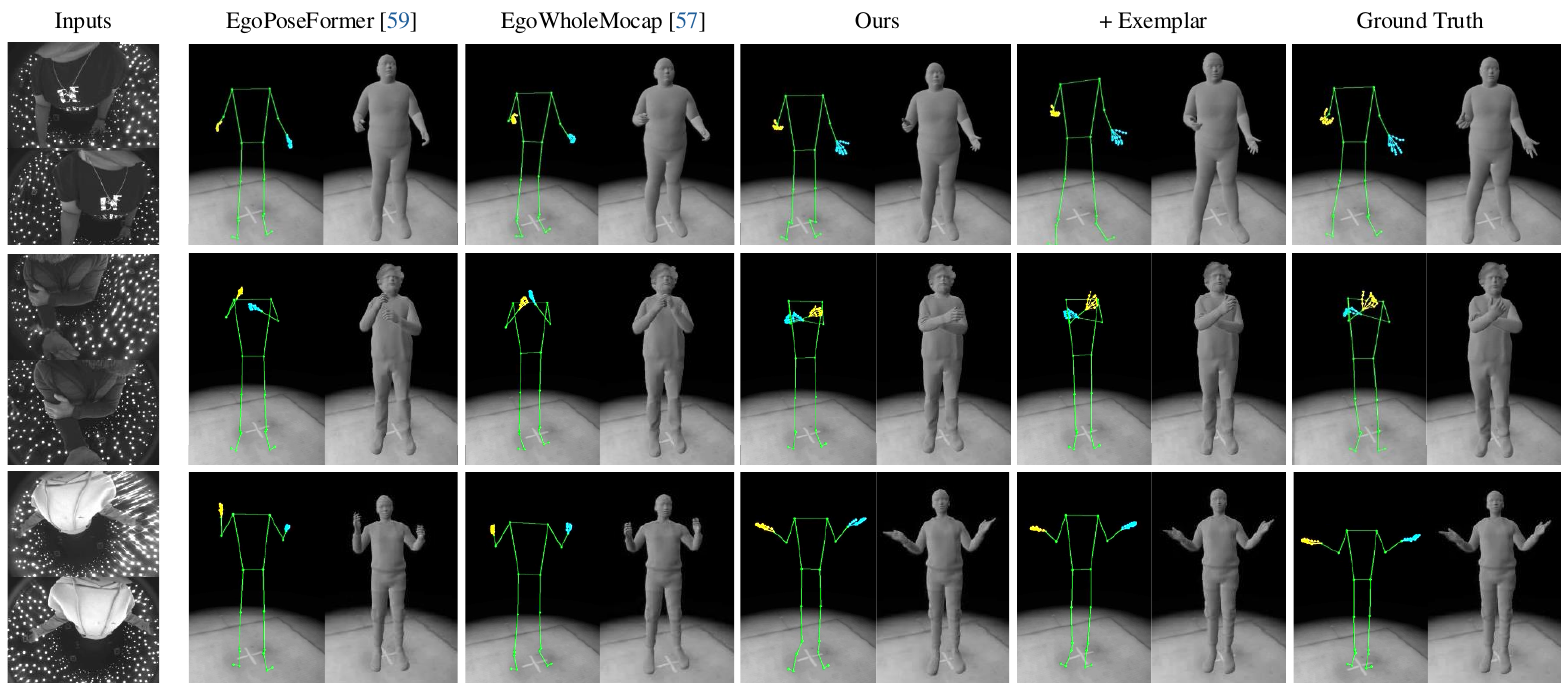}
\vspace{-0.6\baselineskip}
\caption{\textbf{Qualitative comparisons on the ColossusEgo dataset.}
While our framework estimates 3D keypoints, we also employ inverse kinematics with per-identity meshes for more effective visual comparisons (refer to the supplementary material for details).
Our method estimates significantly more accurate and natural motions compared to the existing state-of-the-art methods~\cite{yang2024egoposeformer, wang2024egocentric}.
The additional exemplar-based identity prior further enhances motion accuracy.} 
\label{fig:baseline_comparisons}
\vspace{-2\baselineskip}
\end{center}
\end{figure*}

\keyword{ColossusEgo.}
We have collected a large-scale real dataset consisting of \emph{over 2.8M frames of 500 identities} performing diverse social motions while wearing head-mounted stereo cameras.
To the best of our knowledge, this is the largest \emph{real} image dataset for egocentric first-person pose and motion estimation.
To obtain accurate 3D pose annotations, we use a multi-view capture system with 200 calibrated cameras.
We apply 2D keypoint detection from highly dense viewpoints, followed by triangulation, to annotate precise 3D whole-body keypoints (see the ground truth samples in Fig.~\ref{fig:baseline_comparisons}).
For our experiments, we randomly sample captures from 20 identities for validation and 30 for testing, with the remaining captures used for training.

\keyword{UnrealEgo~\cite{akada2022unrealego,akada20243d}.}
UnrealEgo1~\cite{akada2022unrealego} and UnrealEgo2~\cite{akada2022unrealego} are synthetic datasets created by rendering RenderPeople~\cite{renderpp} 3D human models performing Mixamo~\cite{mixamo} motions.
Although originally proposed for egocentric body-only pose estimation~\cite{wang2021estimating, cuevas2024simpleego}, these datasets provide auxiliary hand pose annotations and temporal sequences, making them suitable for our validation.
For our experiments, we use samples from both UnrealEgo1 and UnrealEgo2, while filtering out sequences shorter than 2 seconds.
We randomly sample 200 sequences for validation and 300 sequences for testing, with the remaining sequences used for training\footnote{The official test set of UnrealEgo2~\cite{akada20243d} does not contain hand annotations.}.
Note that we do not use this dataset for identity-aware motion estimation experiments, as its ground truth motions are not identity-dependent.


\input{tab/main_results}

\subsection{Baselines and Evaluation Metrics}
\label{subsubsec:baseline}
\noindent \textbf{Baselines.}
We consider the two most recent state-of-the-art methods for body pose or motion estimation from down-facing egocentric cameras: EgoWholeMocap~\cite{wang2024egocentric} and EgoPoseFormer~\cite{yang2024egoposeformer}.
EgoWholeMocap~\cite{wang2024egocentric} is the most relevant baseline, as it is the first egocentric \emph{whole-body} motion estimation method.
However, since it was originally designed for monocular egocentric image inputs, we modified its reverse motion diffusion process to adapt to stereo-based pose estimates for fair comparisons.
EgoPoseFormer~\cite{yang2024egoposeformer} is the most recently proposed egocentric pose estimation method, but it only estimates body keypoints. To ensure a fair comparison, we extended its framework to predict whole-body keypoints.
For more details on the baseline modifications, please refer to the supplementary material.

\noindent \textbf{Temporal inference.}
EgoPoseFormer~\cite{yang2024egoposeformer} and our method inherently generalize to arbitrary-length motions due to the use of a frame-based model and a temporal model invariant to the input sequence lengths, respectively.
In contrast, EgoWholeMocap~\cite{wang2024egocentric} assumes a fixed motion length of $T=196$.
Thus, for fair comparisons, we evaluate all models on test sequences adjusted to lengths that are multiples of 196.
We later show that, despite being trained on motion segments of $T=50$, our model seamlessly generalizes to longer motions and outperforms EgoWholeMocap.

\noindent \textbf{Evaluation metrics.}
We use \emph{Mean Per Joint Position Error} (MPJPE) and \emph{Procrustes-Aligned Mean Per Joint Position Error} (PA-MPJPE), which are commonly used to evaluate the accuracy of human motion estimation~\cite{akada20243d,wang2024egocentric,wang2021estimating,rhodin2016egocap}.
We also report \emph{Foot Skate}, which measures foot sliding distance~\cite{tripathi2025humos}, and \emph{Bone Err.}, which is the L2 distance between the predicted and ground truth bone lengths.
All metrics are reported in millimeters.
For the diffusion-based methods (ours and EgoWholeMocap~\cite{wang2024egocentric}), we follow~\cite{wang2024egocentric} and report the average scores of five evaluations.
%
\vspace{-0.4\baselineskip}
\begin{figure*}[!t]
\begin{center}
\includegraphics[width=0.97\textwidth]{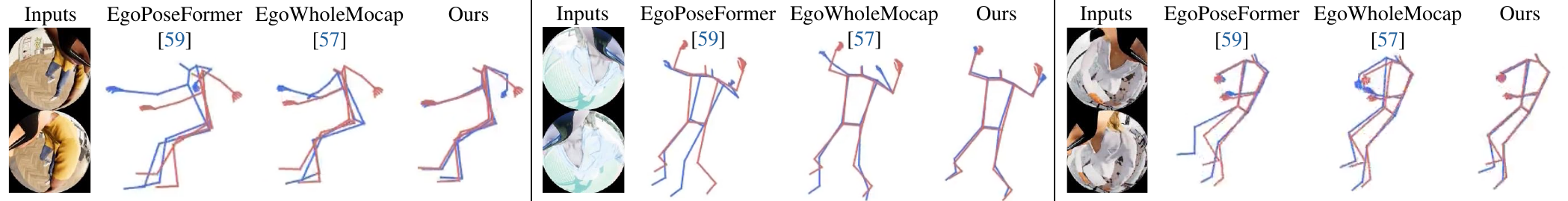}
\vspace{-0.7\baselineskip}
\caption{\textbf{Qualitative comparisons on the UnrealEgo dataset~\cite{akada2022unrealego,akada20243d}.} Red represents the ground truth skeleton, while blue represents the predicted skeleton. Our method estimates more accurate motions compared to the existing baselines~\cite{wang2024egocentric, yang2024egoposeformer}.}
\label{fig:unrealego_comparisons}
\vspace{-\baselineskip}
\end{center}
\end{figure*}

\input{tab/ablation_results}

\vspace{0.3\baselineskip}
\subsection{Egocentric Whole-Body Motion Estimation}
\label{subsec:egocentric_tracking_experiments}
In Tab.~\ref{table:main_results}, we report the quantitative comparison results for egocentric whole-body motion estimation, where our method outperforms the baselines across all metrics on both datasets.
Note that EgoWholeMocap~\cite{wang2024egocentric} performs motion refinement using an \emph{unconditional} motion prior.
As a result, we observe that the output motions are less aligned with the input egocentric observations when the initial estimates are suboptimal.
While EgoPoseFormer~\cite{yang2024egoposeformer} performs direct keypoint estimation similar to ours, it estimates poses on a per-frame basis without utilizing temporal context.
For qualitative comparisons, please refer to Fig.~\ref{fig:baseline_comparisons}-\ref{fig:unrealego_comparisons} and the supplementary video, where motions estimated by our method appear \emph{significantly more accurate and natural}.


\subsection{Identity-Aware Motion Estimation}
\label{subsec:identity_prior_experiments}
\vspace{-0.2\baselineskip}
We now investigate the effectiveness of our exemplar-based identity conditioning method for estimating identity-aware motion.
For the baselines, we consider settings where the identity condition is available in the form of height, shape parameters, and bone lengths.
In Tab.~\ref{subtable:baseline_comparisons}, our exemplar-based identity conditioning is the most effective among these baselines.
Note that $\mathrm{Exemplar}$ denotes our main identity conditioning method based on 10 example poses of the target identity predicted from monocular images, while $\mathrm{Exemplar}^{\dag}$ represents a variant that utilizes 10 ground truth poses, e.g., obtained through a multi-view capture process~\cite{habermann2021real}.
Also refer to our qualitative results in Fig.~\ref{fig:baseline_comparisons}, where the exemplar-based identity conditioning effectively reduce motion ambiguities, e.g., leading to motions that better capture the person’s lower body posture style in the first row of Fig.~\ref{fig:baseline_comparisons}.
To the best of our knowledge, this is the \emph{first} study to analyze the effectiveness of identity priors in egocentric motion estimation.
%



\subsection{Ablation Study}
\label{subsec:ablation_study}
In Tab.~\ref{table:ablation_study}, we present our ablation study results to investigate the effectiveness of each of the proposed modules.

\noindent \textbf{Regression vs. diffusion (Row A).} 
\textit{No Diffusion} represents a variant of our method where motion estimation is performed via regression instead of denoising diffusion.
%
%
Our method outperforms this variant, which aligns with the observations from existing diffusion-based pose or motion estimation works~\cite{stathopoulos2024score, holmquist2023diffpose, zhou2023diff3dhpe, gong2023diffpose, feng2023diffpose, zhang2023probabilistic, yi2024estimating, wang2024egocentric}.

\noindent \textbf{Cascaded body-hand estimation (Rows B-C).}
\textit{Sep. Body-Hand} and \textit{Joint Body-Hand} represent our method variants in which body and hand keypoints are separately estimated and whole-body keypoints are jointly estimated, respectively.
Compared to these variants, our cascaded approach yields better results due to the advantages discussed in Sec.~\ref{subsubsec:decomposed_body_hand_diffusion}.

\noindent \textbf{Temporal network architecture (Row D).}
\textit{Autoregressive} is our method variant using autoregressive modeling, which could serve as an alternative for estimating arbitrary-length sequences.
However, autoregressive models have some limitations, such as exposure bias from teacher forcing~\cite{medsker2001recurrent}, due to the direct reliance on previous estimation outputs.
Our proposed model outperforms this variant, validating our design choice.

\noindent \textbf{Diffusion distillation (Row E).}
\textit{No Diffusion Distill.} is our method variant where a one-step diffusion model is directly trained without diffusion distillation.
Our distilled model (Row F) achieves better results.
For reference, we also report the results of the multi-step teacher diffusion model in Row G.
While our one-step diffusion model yields the best results for real-time tracking, the multi-step diffusion model still provides superior performance, offering an alternative for applications without efficiency constraints.

\keyword{Time comparisons.}
We note that the inference time of our framework is 32 $ms$ and 274 $ms$ with and without distillation, respectively, on a single A100 GPU.
The inference time of the existing SotA baseline (EgoWholeMocap~\cite{wang2024egocentric}) is 2576 $ms$ due to the iterative post-processing steps.





%% file: tab/main_results.tex
\begin{table*}[!t]
\centering
\footnotesize{
\renewcommand{\arraystretch}{1.0}
\caption{\textbf{Quantitative comparisons on egocentric whole-body motion estimation.}}
\vspace{-0.8\baselineskip}
\label{table:main_results}

\begin{subtable}[t]{\textwidth}
\caption{\textbf{Comparison results on the ColossusEgo dataset.} In Rows A-C, our approach outperforms the existing SotA egocentric pose and motion estimation methods~\cite{yang2024egoposeformer,wang2024egocentric} in all metrics. 
In Rows D-H, our exemplar-based identity priors achieve higher performance improvements compared to other identity priors.
$\mathrm{Exemplar}$ and $\mathrm{Exemplar}^{\dag}$ denote our identity-conditioning method with the estimated and the ground truth example 3D poses, respectively. 
}  
\begin{tabularx}{\textwidth}{>{\centering}m{0.3cm}|>{\centering}m{2.8cm}|Y|Y|Y|Y|Y|Y}
\toprule
&Method & $\textrm{MPJPE}_{\,\textrm{Body}}$ & $\textrm{PA-MPJPE}_{\,\textrm{Body}}$ & $\textrm{MPJPE}_{\,\textrm{Hand}}$ & $\textrm{PA-MPJPE}_{\,\textrm{Hand}}$ & Bone Err. & Foot Skate \\
\midrule
A&EgoPoseFormer~\cite{yang2024egoposeformer} & 64.01 & 49.62 & 33.29 & 15.23 & 13.07 & 1.63 \\
B&EgoWholeMocap~\cite{wang2024egocentric} & 62.49 & 43.26 & 25.67 & 12.83 & 10.78 & 0.46 \\
C&\textbf{\textsc{Rewind} (Ours)} & \textbf{53.83} & \textbf{41.42} & \textbf{21.18} & \textbf{10.21} & \textbf{9.78} & \textbf{0.21} \\
\midrule
D&+ Height & 51.98 & 40.39 & 21.10 & 9.80 & 9.24 & 0.21 \\
E&+ Shape Parameters & 51.40 & 39.43 & 21.17 & 10.10 & 7.31 & 0.20 \\
F&+ Bone Lengths & 49.74 & 39.40 & 19.81 & 9.85 & 6.19 & 0.21 \\
G&\textbf{+ $\textrm{Exemplar}$ (Ours)} & \textbf{48.45} & \textbf{33.15} & \textbf{19.20} & \textbf{9.03} & \textbf{5.86} & \textbf{0.17} \\
H&\textcolor{gray}{+ $\textrm{Exemplar}^{\dag}$ (Ours)} & \textcolor{gray}{38.99} & \textcolor{gray}{28.52} & \textcolor{gray}{17.33} & \textcolor{gray}{8.34} & \textcolor{gray}{3.47} & \textcolor{gray}{0.18} \\
\bottomrule
\end{tabularx}
\label{subtable:baseline_comparisons}
\end{subtable}

\vspace{0.3\baselineskip}
\begin{subtable}[t]{\textwidth}
\caption{\textbf{Comparison results on the UnrealEgo dataset~\cite{akada2022unrealego,akada20243d}.} Ours outperforms the baselines across all metrics. Note that the \textit{Foot Skate} metric is not considered for this dataset, as the motions are defined in a camera-centric coordinate system.}

\begin{tabularx}{\textwidth}{>{\centering}m{0.3cm}|>{\centering}m{2.8cm}|Y|Y|Y|Y|Y}
\toprule
&Method & $\textrm{MPJPE}_{\,\textrm{Body}}$ & $\textrm{PA-MPJPE}_{\,\textrm{Body}}$ & $\textrm{MPJPE}_{\,\textrm{Hand}}$ & $\textrm{PA-MPJPE}_{\,\textrm{Hand}}$ & Bone Err.\\
\midrule
A&EgoPoseFormer~\cite{yang2024egoposeformer} & 53.74 & 41.83 & 25.19 & 11.52 & 8.61 \\
B&EgoWholeMocap~\cite{wang2024egocentric} & 49.10 & 39.25 & 25.07 & 10.59 & 9.01 \\
C&\textbf{\textsc{Rewind} (Ours)} & \textbf{37.23} & \textbf{28.04} & \textbf{20.45} & \textbf{9.04} & \textbf{6.22} \\
\bottomrule
\end{tabularx}
\label{subtable:identity_conditioning_experiments}
\end{subtable}
}
\vspace{-0.5\baselineskip}
\end{table*}

%% file: tab/ablation_results.tex
\begin{table*}[!t]
\centering
\footnotesize{
\renewcommand{\arraystretch}{1.0}
\caption{\textbf{Ablation study results on the ColossusEgo dataset (Sec.~\ref{subsec:ablation_study}).} Our method outperforms its variants across all metrics.}
\label{table:ablation_study}
\vspace{-0.5\baselineskip}
\begin{tabularx}{\textwidth}{>{\centering}m{0.3cm}|>{\centering}m{2.9cm}|Y|Y|Y|Y|Y|Y}
\toprule
&Method & $\textrm{MPJPE}_{\,\textrm{Body}}$ & $\textrm{PA-MPJPE}_{\,\textrm{Body}}$ & $\textrm{MPJPE}_{\,\textrm{Hand}}$ & $\textrm{PA-MPJPE}_{\,\textrm{Hand}}$ & Bone Err. & Foot Skate \\
\midrule
A & No Diffusion & 57.34 & 44.26 & 27.04 & 14.99 & 10.83 & \textbf{0.20} \\
B &Sep. Body-Hand & - & - & 23.07 & 11.23 & - & - \\
C &Joint Body-Hand & 55.09 & 42.47 & 24.02 & 11.34 & 10.73 & 0.21 \\
D &Autoregressive & 58.14 & 44.33 & 24.19 & 10.70 & 10.72 & 0.21 \\
E & No Diffusion Distill. & 56.12 & 43.85 & 21.34 & 10.33 & 11.05 & \textbf{0.20} \\
F &\textbf{\textsc{Rewind} (Ours)} & \textbf{53.83} & \textbf{41.42} & \textbf{21.18} & \textbf{10.21} & \textbf{9.78} & 0.21 \\
G &\textcolor{gray}{\textsc{Rewind} (Multi-Step)} & \textcolor{gray}{46.18} & \textcolor{gray}{35.26} & \textcolor{gray}{20.85} & \textcolor{gray}{9.43} & \textcolor{gray}{4.91} & \textcolor{gray}{0.21} \\
\bottomrule
\end{tabularx}
\vspace{-0.5\baselineskip}
}
\vspace{-0.5\baselineskip}
\end{table*}

%% file: sec/5_conclusion.tex
\section{Conclusion}
\label{sec:conclusion}

We introduced a real-time, fully causal framework that enables high-quality whole-body motion estimation from egocentric images. 
To this end, we proposed (1) cascaded denoising diffusion, (2) a  causal relative-temporal Transformer trained with diffusion distillation, and optionally, (3) exemplar-based identity conditioning.
We empirically showed that ours leads to more accurate and natural motions compared to the competitive baselines.

\noindent \textbf{Limitations.}
Although our method outperforms existing state-of-the-art methods, we observed that a small portion of the reconstructed motions leads to self-penetrations. 
Investigating effective methods to avoid self-penetrations in egocentric human motion estimation would be an interesting direction for future work.

\vspace{3\baselineskip}
\noindent \textbf{Acknowledgements.}
Jihyun Lee thanks Soyong Shin (CMU) and Jiye Lee (SNU) for the insightful discussions on motion diffusion models. She also thanks Carter Tiernan (Codec Avatars Lab, Meta) for the help with the ColossusEgo dataset.
T-K. Kim was supported by the NST grant (CRC 21011, MSIT), IITP grants (RS-2023-00228996,
RS-2024-00459749, MSIT) and the KOCCA grant (RS-2024-00442308, MCST).
M. Sung was supported by the NRF grant (RS-2023-00209723) and IITP grants (RS-2022-II220594, RS-2023-00227592, RS-2024-00399817), funded by the Korean government (MSIT).

%
%
%
%
%

%
%
%

%



%% file: sec/6_suppl.tex
\clearpage
\setcounter{page}{1}
\maketitlesupplementary

\setcounter{figure}{0}
\setcounter{table}{0}
\counterwithin{figure}{section}
\counterwithin{table}{section}

\renewcommand{\thesection}{S}
\renewcommand{\thetable}{S\arabic{table}}
\renewcommand{\thefigure}{S\arabic{figure}}

\subsection{Video Results}
\label{sec:video_results}
The video results of our main qualitative comparisons (Fig.~\textcolor{RoyalBlue}{3}-\textcolor{RoyalBlue}{4} in the main paper) are available at \url{https://youtu.be/sMEGyQKHr8c}.
In the video, our method is shown to estimate significantly more accurate and natural motions compared to the existing baselines (EgoWholeMocap~\cite{wang2024egocentric} and EgoPoseFormer~\cite{yang2024egoposeformer}).
%

\subsection{Results with Varying Numbers of Example Poses}
\label{sec:additional_ablation_study}
In Table~\ref{table:supp_ablation_study}, we present additional results on exemplar-based identity conditioning with varying numbers of example poses.
For our main experiments (Sec.~\textcolor{RoyalBlue}{4.4}), we use 10 example poses ($N_{\textit{ex}} = 10$).
We observed that using fewer than 10 example poses ($N_{\textit{ex}}=5$) leads to a degradation in motion estimation quality.
Conversely, significantly increasing the number of example poses ($N_{\textit{ex}}=25$) slightly improves body motion accuracy, but does not enhance hand motion accuracy.
Based on these findings, we selected $N_{\textit{ex}}=10$ for our main experiments, as it provides a good balance between motion accuracy and the ease of example pose acquisition.

\input{tab/supp_ablation_results}




\subsection{Implementation Details}
\label{sec:implementation_details}

In this section, we provide additional implementation details of our whole-body motion estimation model.
\subsubsection{Input Encoding}
Recall that our network takes as input a sequence of egocentric observations $\Phi^{1:T}$, consisting of stereo images and camera poses, along with a sequence of diffused keypoints $\tilde{\mathbf{J}}^{1:T}_{k}$ and the corresponding diffusion time $k$.
We first describe how each of the conditioning inputs is encoded.

\noindent \textbf{Egocentric images.}
We first estimate 2D keypoints from the egocentric images to encode the geometric information.
In particular, we use an EfficientNet-based encoder~\cite{tan2019efficientnet} and a CNN-based decoder~\cite{lecun2015deep} to estimate 2D heatmaps.
Our encoder consists of four stacks of EfficientNet~\cite{tan2019efficientnet} blocks, each containing three mobile inverted bottlenecks~\cite{sandler2018mobilenetv2} with width multipliers of $[16, 24, 40]$ and depth multipliers of $[1, 2, 2]$, followed by Hard Swish~\cite{howard2019searching} activation.
%
%
Our decoder consists of three stacks of convolutional blocks, each containing two 2D convolutional layers, followed by batch normalization~\cite{ioffe2015batch} and ReLU~\cite{agarap2018deep} activation.
%

\noindent \textbf{Camera poses.}
Recall that each camera pose corresponding to viewpoint $v$ is represented by the camera rotation $\mathbf{R}_{v} \in \mathbb{R}^{3 \times 3}$ and translation $\mathbf{t}_{v} \in \mathbb{R}^{3 \times 1}$.
We first convert the camera rotation to a 6D rotation representation~\cite{zhou2019continuity} and concatenate it with the camera translation vector.
We then apply a two-layer MLP, with output feature dimensions set to 256 and 512 for the student and teacher models, respectively.
We use Swish~\cite{ramachandran2017searching} activation for the first layer, while the second layer has no activation.

\noindent \textbf{Diffusion timestep.}
We encode the input diffusion timestep based on the sinusoidal functions, as proposed in \cite{ho2020denoising,gong2023diffpose}.
We then apply a two-layer MLP with the same network architecture as the camera pose encoder.

\noindent \textbf{Upper body poses.}
Our hand model additionally uses 3D upper body keypoints as conditioning inputs.
We flatten the upper body keypoints and apply a two-layer MLP with the same network architecture as the camera pose encoder.
Note that we use the ground truth upper body keypoints during training, while during testing, we use the keypoints predicted by the body model.

\subsubsection{Frame Feature Extraction}
We now extract frame-wise features by aggregating the conditioning input features.
In particular, we concatenate the estimated stereo 2D keypoints with the confidence scores to the corresponding diffused keypoints $\tilde{\mathbf{J}}^{t}_{k}$ at each timestep $t$.
We additionally concatenate the features of (1) stereo camera poses, (2) the diffusion timestep, and (3) an upper body pose (only for the hand model) to the corresponding diffused keypoints.
We then apply Graph Transformer blocks~\cite{gong2023diffpose}, which consist of graph convolution~\cite{defferrard2016convolutional} and self-attention~\cite{vaswani2017attention} layers, on the human skeletal graph.
For the teacher network, we use three Graph Transformer blocks, with feature dimensions set to 512 and the number of attention heads set to 4.
For the student network, we use a single block with a feature dimension of 256 and 2 attention heads to enable faster inference.
Note that we use a linear layer to estimate poses from these intermediate frame-based features to incorporate auxiliary reconstruction loss (to be explained in Sec.~\ref{subsubsec:network_training}).

\subsubsection{Temporal Feature Extraction}
Given the frame-based features extracted for each timestep $t$, we apply our Causal Relative-Temporal Transformer (Sec.~\textcolor{RoyalBlue}{3.2}) to extract temporal features.
For the teacher model, we use three relative-temporal attention layers with 4 attention heads and a window size of 20.
For the student model, we use a single relative-temporal attention layer with 2 attention heads with a window size of 8.
We set the output feature dimensions to 512 and 256 for the teacher and student models, respectively.
Finally, we apply a linear layer to map the output temporal features to the sequence of whole-body keypoints.

\subsubsection{Network training.}
\label{subsubsec:network_training}
We follow DDPM~\cite{ho2020denoising} for training our diffusion model.
For the teacher network, we diffuse the ground truth keypoints with a randomly sampled diffusion timestep $k \in [1, K]$ and feed them as inputs to the network.
For the student network, the diffusion timestep is set to the maximum value $k = K$ to enable single-step sampling.
For noise scheduling, we use cosine scheduling from $\beta_{1} = 0.0001$ to $\beta_{K} = 0.02$ with the maximum diffusion timestep set as $K = 1000$ (refer to \cite{ho2020denoising} for details on the noise scheduling hyperparameter $\beta_{k}$).

We train our diffusion network for 2M steps using an Adam optimizer with a learning rate of
$5\times10^{-5}$.
We use a single batch consisting of $T=50$ consecutive frames for training, though our network can inherently generalize to motions longer than the training sequences due to the proposed architecture~(Sec.~\textcolor{RoyalBlue}{3.2}). 
For the training loss, we mainly adopt the loss function from MDM~\cite{tevet2022human}, which includes: (1) $\mathcal{L}_{\textit{simple}}$, the L2 distance between the predicted and ground truth motion signals at $k=0$, (2) $\mathcal{L}_{\textit{vel}}$, the L2 distance between the predicted and ground truth motion velocities, and (3) $\mathcal{L}_{\textit{foot}}$, which regularizes the slided foot keypoints (refer to \cite{tevet2022human} for computation details).
We additionally use $\mathcal{L}_{\textit{frame}}$, an auxiliary L2 loss between the poses predicted from intermediate frame-based features and the ground truth poses. 
Our final loss function, $\mathcal{L}_{\textit{total}}$, is defined as:
\begin{equation}
\mathcal{L}_{\textit{total}} = \mathcal{L}_{\textit{simple}} + \lambda_{\textit{vel}}\,  \mathcal{L}_{\textit{vel}} + \lambda_{\textit{foot}}\, \mathcal{L}_{\textit{foot}} + \lambda_{\textit{frame}}\, \mathcal{L}_{\textit{frame}}.
\end{equation}
For $\lambda_{\textit{vel}}$, $\lambda_{\textit{foot}}$ and $\lambda_{\textit{frame}}$, we initially use values of 300, 100, and 1, respectively.
However, we observe that the loss terms involving motion velocities ($\mathcal{L}_{\textit{vel}}$ and $\mathcal{L}_{\textit{foot}}$) converge to very small values in the later stages of training.
Thus, we increase the values for $\lambda_{\textit{vel}}$ and $\lambda_{\textit{foot}}$ to 4K and 20K, respectively, in the last 10K training steps.
Note that, for training the student model, we additionally use the distillation loss $\mathcal{L}_{\textit{distill}}$ (Sec.~\textcolor{RoyalBlue}{3.3}) with the weighting hyperparameter $\lambda_{\textit{distill}}$ set to 1.

\paragraph{Network inference.}
We use DDIM~\cite{song2021denoising} for network inference, with the number of sampling steps set to 10 for the teacher network and 1 for the student network.

\subsection{Details on Inverse Kinematics}
\label{subsec:inverse_kinematics}
To use our motion tracking results for driving meshes or avatars (e.g., through linear blend skinning), we optionally perform inverse kinematics to convert the estimated 3D keypoints to joint rotations.
To this end, we train a simple inverse kinematics network that takes as inputs the 3D whole-body keypoints along with the stereo camera translations (for estimating head poses) per frame and outputs joint rotations.
We build our network upon the Graph Transformer~\cite{gong2023diffpose}, similar to the frame-based feature extraction module in our main diffusion model.
We use five Graph Transformer blocks~\cite{gong2023diffpose}, with output feature dimensions and the number of attention heads set to 512 and 4, respectively.  
After the last layer, we use a linear layer to map the final features to the joint rotations in a 6D rotation representation~\cite{zhou2019continuity}.
For network training, we use L2 loss between the predicted and ground truth joint rotations.
We train the network with an Adam optimizer and a learning rate of $5\times10^{-5}$.

\subsection{Details on Example Pose Estimation}
To estimate 3D example poses of the target identity from monocular images, we perform parametric body model fitting to the pseudo ground truth 2D keypoints and depth estimated by Sapiens~\cite{khirodkar2025sapiens}, an off-the-shelf foundational human model.
In particular, we fit the parametric body model using the loss $\mathcal{L}_{\textit{opt}}$ defined as:

\vspace{-\baselineskip}
\begin{equation}
\mathcal{L}_{\textit{opt}} = \mathcal{L}_{\textit{2D}} + \lambda_{\textit{depth}}\, \mathcal{L}_{\textit{depth}} + \lambda_{\textit{reg}}\, \mathcal{L}_{\textit{reg}} + \lambda_{\textit{height}}\, \mathcal{L}_{\textit{height}}.
\end{equation}

\noindent where $\mathcal{L}_{\textit{2D}}$ is the L2 loss between the 2D projection of the predicted 3D keypoints and the pseudo ground truth 2D keypoints. 
$\mathcal{L}_{\textit{depth}}$ is the L2 loss between the predicted and the pseudo ground truth depth maps.
$\mathcal{L}_{\textit{reg}}$ is the L2 loss between the current body model parameters and the mean body model parameters in the training set, penalizing deviations from the mean parameters.
We also incorporate $\mathcal{L}_{\textit{height}}$, which measures the L2 distance between the predicted and the ground truth height of the target identity to reduce the scale ambiguity.
We set $\lambda_{\textit{depth}}$, $\lambda_{\textit{reg}}$, and $\lambda_{\textit{height}}$ to 100, 300, and 1, respectively.
We perform 3K optimization iterations using the AdamW optimizer~\cite{loshchilov2017decoupled} with an initial learning rate of $5 \times 10^{-3}$.
The learning rate is decayed by 0.023\% after each optimization iteration.
%



\subsubsection{Details on Baseline Comparisons}

\paragraph{EgoWholeMocap~\cite{wang2024egocentric}.}
EgoWholeMocap is the first egocentric whole-body motion estimation method, making it the most relevant baseline for our work.
It estimates frame-based 3D poses through 2.5D heatmap estimation and undistortion using the camera parameters, followed by temporal refinement with an unconditional motion diffusion model, where its DDPM~\cite{ho2020denoising}-based motion sampling is guided by the initial 3D poses and their uncertainty scores.
In particular, given the clean motion signal $\hat{\textbf{x}}_{0}$ estimated by the diffusion model at each diffusion timestep $k$, it defines the mean of the Gaussian distribution for sampling $x_{k-1}$ as:

\begin{equation}
\hat{\textbf{x}}_{0} + \textbf{w}(\textbf{x}_{e} - \hat{\textbf{x}}_{0}),
\label{eq:egowholemocap_obj}
\end{equation}

\noindent where $\textbf{x}_{e}$ is a sequence of initially estimated whole-body poses, and $\mathbf{w}$ is a weighting vector computed from their uncertainty scores (refer to Eq.~\textcolor{RoyalBlue}{5} in the original paper~\cite{wang2024egocentric}).

Note that its original method considers monocular image inputs. 
To make a fairer comparison to our method, which uses stereo image inputs, we modify the method to (1) estimate 2.5D heatmaps from each of the input stereo images, (2) convert them to 3D poses using the known camera parameters, and (3) perform diffusion-based motion sampling guided by these stereo initial 3D pose estimates by modifying Eq.~\ref{eq:egowholemocap_obj}:

\begin{equation}
\hat{\textbf{x}}_{0} + \frac{\textbf{w}_{L}}{2}(\textbf{x}_{e_{L}} - \hat{\textbf{x}}_{0}) + \frac{\textbf{w}_{R}}{2}(\textbf{x}_{e_{R}} - \hat{\textbf{x}}_{0}),
\end{equation}

\noindent where $\textbf{x}_{e_{v}}$ and $\textbf{w}_{v}$ are the initial poses and uncertainty scores estimated from the input image of viewpoint $v$.

\paragraph{EgoPoseFormer~\cite{yang2024egoposeformer}.}
EgoPoseFormer is one of the most recently proposed stereo egocentric pose estimation methods. 
However, it was originally designed to estimate body-only keypoints.
To enable comparisons with our method, we modify EgoPoseFormer to estimate whole-body keypoints during both the 2D heatmap and 3D pose estimation stages.
Additionally, we incorporate the input camera poses (which are used in our method) by encoding them with an MLP-based encoder and performing feature concatenation in the 3D pose estimation network, similar to our approach.

%% file: tab/supp_ablation_results.tex
\begin{table}[!h]
\centering
\footnotesize{
\renewcommand{\arraystretch}{1.0}
\caption{\textbf{Results with varying numbers of example poses.} 
Except for the number of example poses ($N_{\textit{ex}}$), we use the same experimental settings as those in $\mathrm{Examplar}^{\dag}$ from Table~\textcolor{RoyalBlue}{1b} in the main paper.
}
\label{table:supp_ablation_study}
\vspace{-0.7\baselineskip}
\begin{tabularx}{0.5\textwidth}{>{\centering}m{0.6cm}|Y|Y|Y|Y}
\toprule
$N_{\textit{ex}}$& $\textrm{MPJPE}_{\,\textrm{Body}}$ & $\textrm{PMPJPE}_{\,\textrm{Body}}$ & $\textrm{MPJPE}_{\,\textrm{Hand}}$ & $\textrm{PMPJPE}_{\,\textrm{Hand}}$\\
\midrule
$5$ & 41.23 & 30.03 & 17.83 & 8.61\\
$10$ & 38.99 & 28.52 & \textbf{17.33} & \textbf{8.34}\\
$25$ & \textbf{37.77} & \textbf{27.91} & 17.77 & 8.58\\
\bottomrule
\end{tabularx}
}
\end{table}